# Feasibility of Using Bandwidth Efficient Modulation to Upgrade the CMS Tracker Readout Optical Links

S. Dris, L. Amaral, K. Gill, R. Grabit, A. Pacheco, D. Ricci, J. Troska, F. Vasey

CERN PH-MIC-OE, Geneva, Switzerland

## Abstract

Plans to upgrade the LHC after approximately 10 years of operation are currently being considered at CERN. A tenfold increase in luminosity delivered to the experiments is envisaged in the so-called Super LHC (SLHC). This will undoubtedly give rise to significantly larger data volumes from the detectors, requiring faster data readout. The possibility of upgrading the CMS Tracker analog readout optical links using a bandwidth efficient digital modulation scheme for deployment in the SLHC has been extensively explored at CERN. Previous theoretical and experimental studies determined the achievable data rate using a system based on Quadrature Amplitude Modulation (QAM) to be ~3-4Gbit/s (assuming no error correction is used and for an error rate of ~$10^{-9}$). In this note we attempt to quantify the feasibility of such an upgrade in terms of hardware implementation complexity, applicability to the high energy physics (HEP) environment, technological feasibility and R&D effort required.

# Glossary

**ADSL** – Assymetric Digital Subscriber Line

**AOH** – Analog OptoHybrid

**APV25** – The 0.25μm Analog Pipeline Voltage readout chip

**ARx12** – 12-channel Analog Optoelectronic Receiver

**ASIC** – Application Specific Integrated Circuit

**BER** – Bit Error Rate

**CERN** – The European Organization for Nuclear Research

**CMOS** – Complementary Metal Oxide Semiconductor

**CMS** – The Compact Muon Solenoid experiment at CERN

**DAC** – Digital to Analog Converter

**DSP** – Digital Signal Processing

**FEC** – Forward Error Correction

**FFT** – Fast Fourier Transform

**HEP** – High Energy Physics

**IC** – Integrated Circuit

**ISI** – InterSymbol Interference

**LHC** – The Large Hadron Collider accelerator

*M*-**PAM** – Multi-level Pulse Amplitude Modulation with *M* levels

*M*-**QAM** – Multi-level Quadrature Amplitude Modulation with *M* levels

**NRZ** – Non-Return to Zero

**OFDM** – Orthogonal Frequency Division Multiplexing

**PAM** – Pulse Amplitude Modulation

**PAPR** – Peak to Average Power Ratio

**QAM** – Quadrature Amplitude Modulation

**R&D** – Research and Development

**SLHC** – Super Large Hadron Collider (the proposed future upgrade to the Large Hadron Collider)

**SCM** – SubCarrier Multiplexing

**SEU** – Single Event Upset

**SNR** – Signal to Noise Ratio

**TTC** – Timing and Trigger Control

# 1 Introduction

The next iteration of the CMS Tracker will be operated in the Super LHC (SLHC) [1] environment and will have to employ significantly faster data readout links due to the increase in luminosity foreseen. The cost of the optoelectronic components represents a large fraction of the CMS Tracker electronics budget. Hence a digital system reusing the existing optoelectronic components while delivering sufficient performance for SLHC operation, could potentially be a cost-effective alternative to a full replacement of the installed links. The first step in the feasibility study of such a conversion has been carried out at CERN and an accurate estimate of the performance that can be achieved has been made in [2-4], where it was shown experimentally that multi-Gbit/s data rates are possible over the current 40 MSamples/s analog optical links by employing techniques similar to those used in ADSL. The concept involves using one or more digitally-modulated sinusoidal carriers in order to make efficient use of the available bandwidth.

The extensive testing and characterization performed on the optoelectronic components described in [2] has facilitated the selection of an appropriate bandwidth efficient modulation scheme for a future upgrade. The laser driver and the receiving amplifier ASICs present bottlenecks in the bandwidth of the Tracker optical link, resulting in an overall 3-dB bandwidth of ~150MHz. Given the excellent noise performance and linearity of the system, it is possible to employ a Quadrature Amplitude Modulation (QAM)-based scheme in order to reach Gbit/s data rates within the available bandwidth. In QAM, a higher frequency sinusoidal carrier is modulated in amplitude and phase to produce data symbols that represent several bits each. The existing optical link components, from the Analog OptoHybrid (AOH) to the Analog Optoelectronic Receiver (ARx12), would be retained in the proposed upgrade, with additional components performing digitization, digital modulation/demodulation and optionally other source encoding functions (Figure 1).

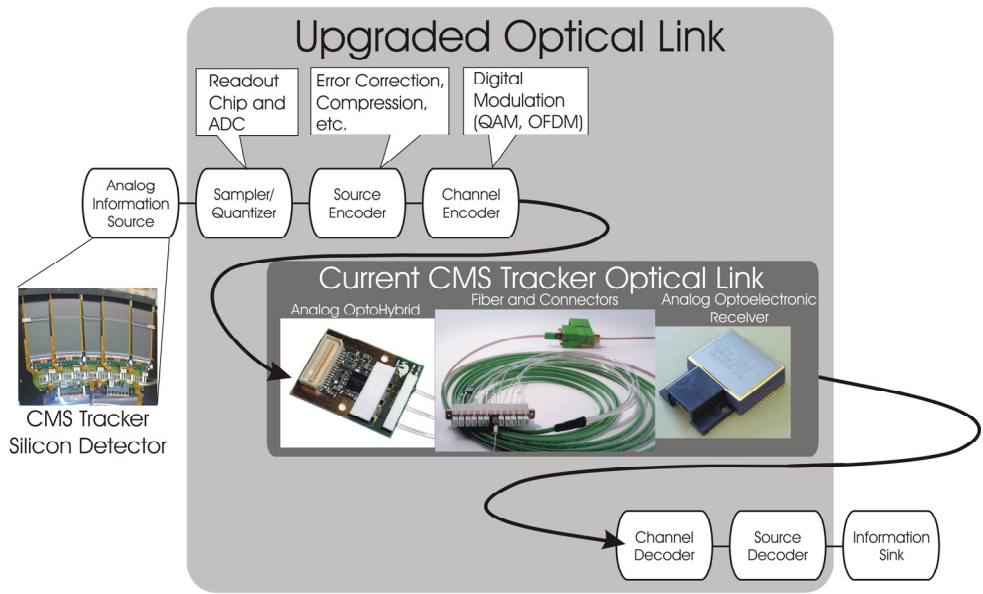

**Figure 1:** The basic building blocks of a digital communication system based on the CMS Tracker analog optical links.

## 1.1 Bandwidth Efficient Modulation: QAM

A brief overview of the basics of bandwidth efficient digital modulation is included in this note, but it is not intended to be exhaustive or complete. For further details, the interested reader should consult any of the multitude of excellent digital communications texts (e.g. [5, 6]).

Modulation schemes based on multi-level signalling improve upon simple binary schemes by making efficient use of a channel's bandwidth. This is achieved by transmitting waveforms corresponding to symbols which convey multiple bits each. In *M*-level digital baseband pulse amplitude modulation (*M*-PAM) for example, each transmitted amplitude level is a symbol corresponding to $\log_2 M$ bits of information[1]. Hence a system that employs 256-PAM has a theoretical maximum bandwidth efficiency of 8bit/s/Hz , compared to 1bit/s/Hz for binary NRZ. However, this eight-fold increase in data rate comes at the price of increased error rate for a given signal to noise ratio (SNR).

A more efficient use of power can be achieved by using passband modulation which employs a sinusoidal carrier, as is the case in QAM. The advantage of this technique over baseband modulation is that the phase of the carrier can be varied *in addition* to the amplitude, providing another dimension over which information bits can be transmitted. As in *M*-PAM, the spectral efficiency of *M*-QAM can theoretically reach $\log_2 M$ bit/s/Hz. However, a QAM system requires a lower SNR per transmitted bit to achieve the same error rate as PAM.

## 1.2 QAM Signal Generation

Combined amplitude and phase modulation is achieved by simultaneously impressing two separate *k*-bit symbols on two quadrature (90° out of phase) sinusoidal carriers (Figure 2). Each branch in Figure 2 is essentially the amplitude modulation (PAM) of the corresponding carrier.

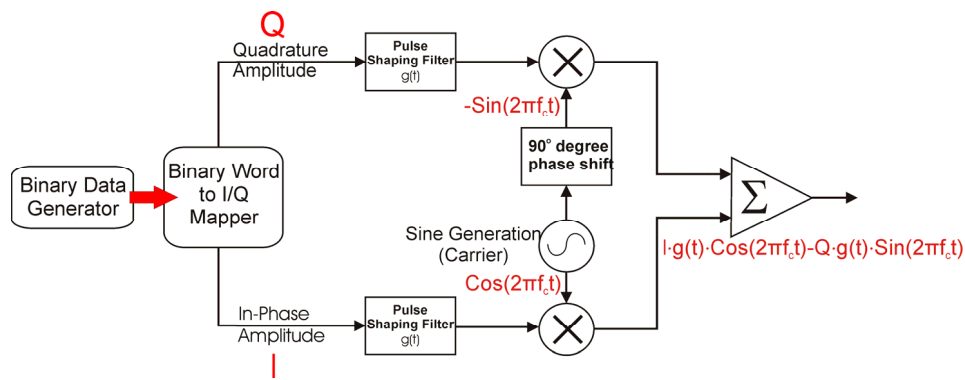

**Figure 2:** Illustration of QAM signal generation

A symbol consisting of 2*k* binary bits is clocked in to the *I/Q* mapper at the required symbol rate. Half of the bits are translated to a quadrature amplitude, *Q*, and half to an in-phase amplitude, *I*. After pulse shaping, *I* and *Q* are modulated on the sin and cos carriers respectively, and summed at the output. The resulting signal waveforms are:

---

[1] Baseband binary schemes such as NRZ can be considered a subset of *M*-PAM, with *M*=2.

$$s(t) = \text{Re}\left[(I + jQ)g(t)e^{j2\pi f_c t}\right] \quad \text{for } 0 \leq t \leq T$$
$$= I \cdot g(t)\cos(2\pi f_c t) - Q \cdot g(t)\sin(2\pi f_c t) \quad (1)$$

Where *g(t)* is the signal pulse and *I* and *Q* are the information-bearing signal amplitudes of the quadrature carriers.

Equivalently, *s(t)* can also be expressed as:

$$s(t) = \text{Re}\left[Ve^{j\theta}g(t)e^{j2\pi f_c t}\right]$$
$$= Ve^{j\theta}g(t)\cos(2\pi f_c t + \theta) \quad (2)$$

Where $V = \sqrt{I^2 + Q^2}$ and $\theta = \tan^{-1}(Q/I)$.

Equation (2) is a more intuitive way of showing that QAM signals are generated by combined amplitude and phase modulation, the resulting waveform simply being a sinusoid of amplitude *V* and phase offset *θ* (with respect to the carrier). The combined number of amplitude levels that can be taken on by *I* and *Q* gives the modulation order, *M*, which in turn determines the number of bits conveyed by each transmitted symbol (=$\log_2 M$). Hence, at each symbol period, an *M*-QAM system transmits any one of *M* available symbols, each denoted by a particular combination of *I* and *Q* values.

## 1.3 Achievable Data Rate Using QAM

In [2-4], fundamental digital communication principles were used to estimate the potential rate in a theoretical future upgrade, taking into account all relevant system parameters such as noise, bandwidth and available transmission power. The current analog link was treated as the communication channel over which digitally modulated signals are to be transmitted. The calculation method has been demonstrated for an uncoded multi-carrier QAM system, though the exact rate achieved in a future system will be implementation-dependent. The method and results are also readily applicable to a single-carrier QAM system.

Experimental evidence of the feasibility of QAM transmission in a Tracker optical link was provided in [2, 3]. Low symbol rate QAM signals with carrier frequencies up to 1GHz were transmitted –one at a time– through an analog optical link with final-version components. This effectively allowed the complete characterization of the link, by evaluating the digital SNR across the entire useable bandwidth of the optical link. The experimental data was fed into the theoretical QAM multi-carrier model, and this has been used to show that, for an uncoded system, 3-4Gbit/s at an error rate of ~$10^{-9}$ are a realistic possibility. The results are shown in Figure 3, where the data rate is plotted against the normalized transmission power injected into the optical link, a system parameter which is implementation-dependent. Curves corresponding to several bit error rates (BER) are given. In general, the transmission power used will depend on whether the future system employs single or multiple carriers, as well as the modulation order (e.g. 16-, 32-, 64-QAM, etc.).

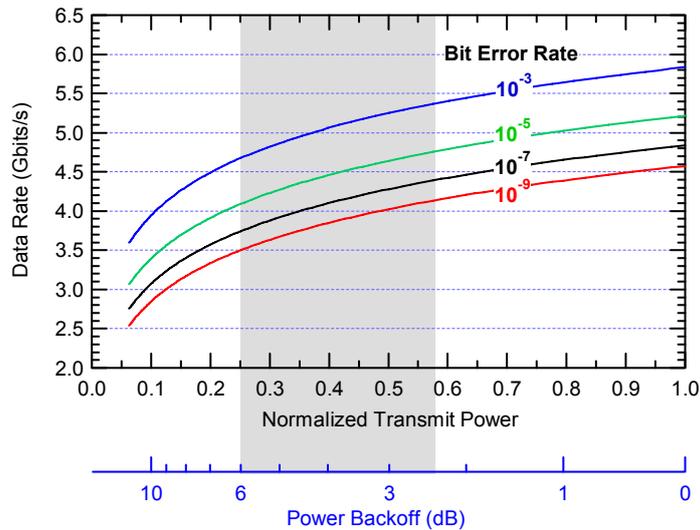

**Figure 3:** Achievable data rate as a function of transmission power (or, equivalently, power 'backoff' from the maximum power that can be injected into the optical links) and BER. The shaded area indicates the likely operating range for a future QAM system.

Section 2 of this note outlines the main considerations and constraints involved in the design of a QAM-based modulation scheme. A hypothetical upgrade scenario is used to illustrate the general parameters of the system, as well as the requirements that must be met in the context of a HEP experiment, and the impact these have on the feasibility of hardware implementation. Section 3 presents a high-speed QAM transceiver ASIC intended for optical link transmission, developed by researchers at Agere Systems. The ASIC's functionality is similar to what would be required for our proposed upgrade and can therefore be used as an example of the performance that can be realistically achieved for a given chip power dissipation. Finally, section 4 summarizes the conclusions of the feasibility study of the QAM-based upgrade.

## 2 QAM Upgrade Case Study: Feasibility of Implementation

### 2.1 Overview

It has been proven that QAM can be used to maximize the data rate in a future optical readout link based on the current CMS Tracker components. However, the applicability of such an upgrade (or any upgrade for that matter) also depends on the suitability of the hardware implementation to the unique requirements of a HEP experiment.

The fundamental trade-off in a digital communication system is that of complexity, power consumption and bandwidth. In general, for a given data rate, a simple modulation scheme (such as binary NRZ) is relatively easy to implement, but requires significant bandwidth. Conversely using bandwidth efficient modulation (e.g. QAM) one can achieve the same data rate and occupy less channel bandwidth, at the expense of more complex transmitters and receivers. The proposed QAM-based upgrade must therefore be evaluated on the basis of whether the increased data rate, complexity and power consumption, as well as the cost of development of additional electronic components, can be justified by the savings in the development and cost of new optoelectronic components.

It will be assumed in this document that the future Tracker readout ASICs will sample the detectors at 40MHz[2]. The optical links will operate at Gbit/s rates and hence multiplexing of the detector data will be required. In the example of Figure 4, the readout ASICs produce data frames consisting of 5-bit samples at 40MS/s. These are then fed into a 16:1 multiplexer to give a 640MS/s (=3.2Gbit/s) data stream. The modulator adds forward error correction (FEC) to the data and produces the 32-QAM (=5bits/symbol) waveform for transmission over the optical link. In the counting room, the waveform is demodulated, digital baseband equalization (EQ) is performed to mitigate the effects of intersymbol interference (ISI) added by the channel, and the FEC-encoded data words are recovered.

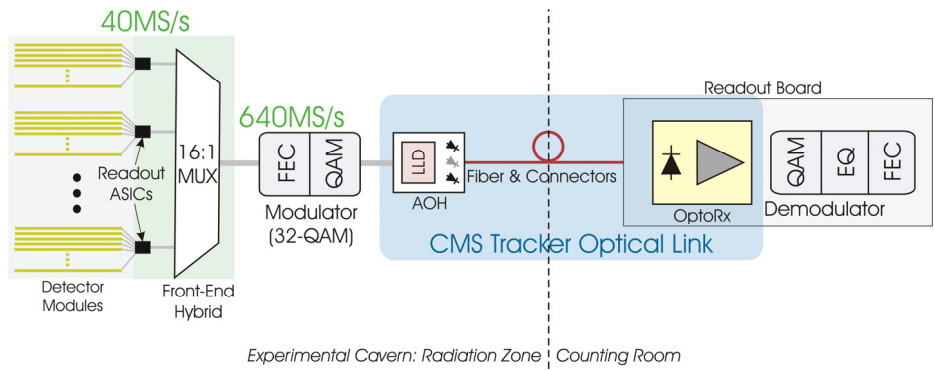

**Figure 4:** A possible configuration for a future QAM-based unidirectional readout system operating at 3.2Gbit/s.

Of primary importance is the transmitter, since it will be placed inside the detector volume (assuming the unidirectional point-to-point link model is retained for the future system). The constraints in the detector front-end are more severe, with the electronics having to endure the harsh radiation environment of a HEP experiment. In addition, the physics performance goals require that the excess material budget inside the experiment is minimized, as this degrades performance. Hence the front-end electronics must satisfy the constraints of low-mass, small size and radiation hardness.

This report therefore focuses on the transmitting side of the link. To summarize, the following issues will need to be addressed for a future system:

- Power consumption
- Radiation hardness
- Low-mass and small size
- Hardware complexity and R&D effort required

## 2.2  Single-Carrier QAM Modulator: Digital vs. Analog Carrier Generation

The basic single-carrier QAM modulator can be realized using digital (Figure 5) or analog (Figure 6) carrier generation. In the fully-digital case, all modulation functions are performed in the digital domain. In order to generate a sinusoidal carrier of sufficient quality, the sampling rate of the system will have to be at least 4 or 8 times that of the carrier frequency. With carrier frequencies expected to reach up to a few hundred MHz, it becomes clear that very high clock speeds would be required in this case, accompanied by higher complexity and power consumption.

---

[2] Other frequencies are also being considered, but this has little impact on the feasibility of the proposed system.

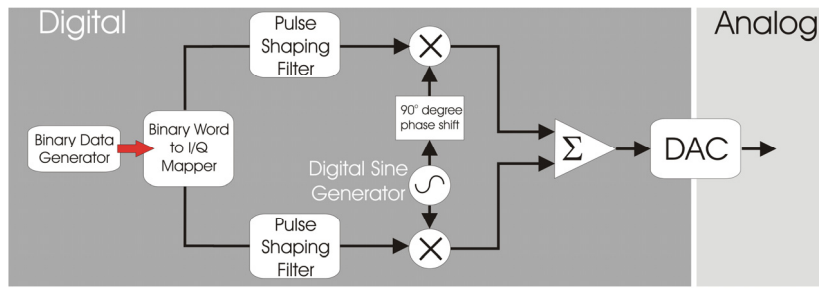

**Figure 5:** The basic building-blocks of the fully-digital QAM modulator.

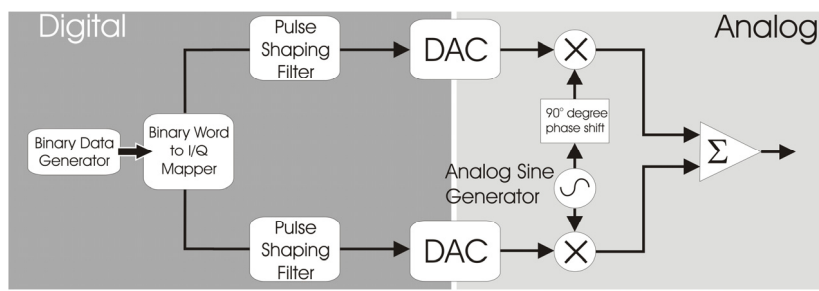

**Figure 6:** The basic building-blocks of the QAM modulator employing analog carrier generation.

A better solution is to generate the carrier in analog form, with DACs to convert the digital baseband QAM signals to analog. In this case, the DACs need only operate at the symbol rate of the modulator and no further upsampling is required (apart from that needed if pulse-shaping is employed, which is common to both schemes shown here).

## 2.3   Single vs. Multi-Carrier

A QAM-based system could employ one carrier as in the modulators shown in the previous section, or multiple carriers as shown in Figure 7. This section describes the main advantages and disadvantages of both techniques. The two systems are illustrated in Figure 8 and Figure 9, where the (hypothetical) QAM signal bandwidths (~300MHz in this example) are superimposed on a frequency response measurement carried out on a CMS Tracker optical link. In the single-carrier case, one high symbol rate carrier takes up the entire (useable) channel spectrum, while a multi-carrier system divides the channel into many, smaller sub-channels, each occupied by a QAM carrier. The two approaches offer different advantages and disadvantages and their suitability for a HEP readout link needs to be evaluated.

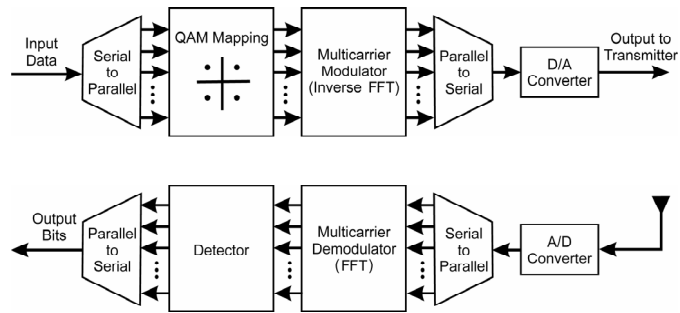

Figure 7: A multi-carrier QAM system.

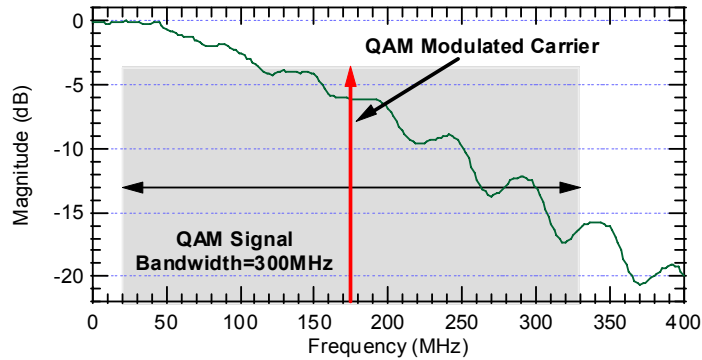

**Figure 8:** Illustrative example of a single, high symbol rate QAM carrier occupying ~300MHz in a frequency-selective channel.

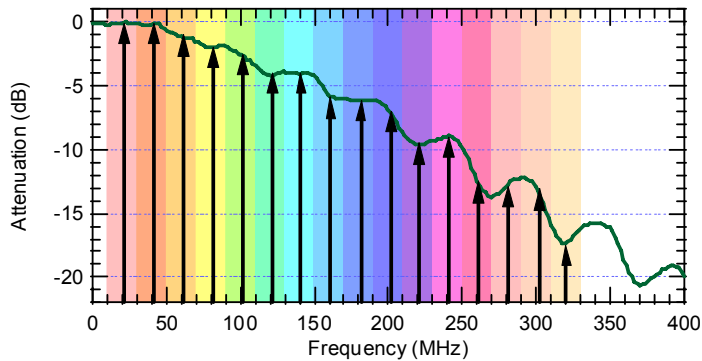

**Figure 9:** Illustrative example of multiple, low symbol rate QAM carriers occupying ~300MHz in a frequency-selective channel.

### 2.3.1 Digital Signal Processing in the Front-End

A multi-carrier modulator employs an inverse fast Fourier transform (FFT) to realize the required orthogonal carriers. The digital to analog conversion occurs at the output of the FFT and the analog signal is transmitted through the channel. It follows that for the multi-carrier case, DSP (in the form of an FFT) will have to be performed by the front-end electronics at a very fast rate. With current electronic technology, it is questionable whether this is possible to achieve while keeping the power consumption within Tracker specifications. For the single-carrier case there is no need to synthesize the carrier frequencies digitally, thus reducing complexity in the front-end.

### 2.3.2 Amplitude/Phase Resolution

The SNR vs. carrier frequency results presented in [2] were derived by measuring the error vector magnitude (EVM) of the QAM carrier under test. EVM is a metric encompassing all channel impairments that corrupt the received signals. It is therefore a universal performance metric for a QAM system and can be used to determine the digital SNR. In this context, the digital SNR includes all effects that contribute to the noisy reception of a symbol, including amplitude noise, jitter, non-linearity, frequency response, etc.

The results showed that if low symbol rate carriers are used, the SNR can reach over 35dB for many of the sub-channels. It follows that as many as 8-10 bits/symbol will be allocated to these high-SNR carriers, thus requiring an extremely high amplitude/phase resolution in the system. If only one high symbol rate carrier is used, it will be susceptible to the full noise spectrum of the channel. The resulting (inferior) SNR will mean that less bits/symbol (~5-6) will be allocated to the single carrier, hence requiring much less resolution than in the multi-carrier case.

### 2.3.3 Peak to Average Power Ratio (PAPR)

Perhaps the most serious disadvantage of a multi-carrier system is that of the peak to average power ratio (PAPR) [2, 7-9]. This can be an order of magnitude higher than in a single-carrier system and requires a significant amount of amplifier backoff (as much as 6-9dB) to avoid saturation that results in symbol errors. Backing-off the transmission power also results in decreased SNR, so a balance between the two needs to be found.

### 2.3.4 Inter-Symbol Interference

One of the disadvantages of the single-carrier system is that of ISI due to the frequency response of the channel. A multi-carrier system is designed such that each carrier occupies a small slice of the available bandwidth that has an almost flat frequency response, effectively eliminating ISI. This is not the case for a carrier occupying the entire channel spectrum and equalization would be required to mitigate the effects of ISI. Hence more complexity is required in the receiver, where very fast equalization (with a sufficiently large number of coefficients) would have to be performed. Given that the frequency response of each analog optical link will not be known a priori, adaptive equalization may be required[3]. This would be computationally expensive at Gbit/s data rates. On the other hand, this may not be a serious drawback if equalizer 'training' needs to be performed only once during the initial setup (i.e. when the link is first assembled). Since each analog optical link has a constant frequency response (i.e. it is not a fading environment as is the case in the wireless world), this may be feasible.

### 2.3.5 Symbol Rate

An upgrade based on one, high symbol rate QAM carrier would be more demanding on the front-end electronics in terms of the requirement for higher clock speeds compared to multiple, low symbol rate carriers. In the case of the Tracker optical link, a very high symbol rate (as high as 500-650MS/s) would be necessary to take full advantage of the available bandwidth of over ~650MHz. This would require a carefully designed high-speed mixed-signal ASIC, including very high-speed DACs (in the GS/s range, assuming oversampling is employed).

---

[3] The frequency response is dominated by the ARx12 amplifier ASIC, and hence might be relatively consistent from link to link, though no measurements have been made to confirm this.

### 2.3.6 Data Rate Adjustment

Finally, an advantage of the multi-carrier case is the ability to adjust the data rate with finer granularity due to the lower system symbol rate. The implication of this is that the data rate can be optimally tailored to each optical link channel, depending on the individual frequency response and noise characteristics. A single QAM carrier operating at, say, 640MS/s would only be adjustable in steps of 640Mbits/s steps (ignoring any encoding). However, adjustability of a link's throughput may not be a required characteristic for a future system, especially since the point-to-point model is very likely to be retained.

### 2.3.7 Summary

Table 1 summarizes the advantages and disadvantages of using single and multiple carrier QAM systems. Clearly, the main advantages of deploying a multi-carrier system is that of lower symbol rates and immunity to ISI (simpler equalization). These, however, come with higher complexity and power consumption due to the need to perform a very fast FFT in the front-end, as well as reduced performance due to high PAPR. The benefits are outweighed by the disadvantages, since multi-carrier systems are normally more appropriate in frequency fading environments where the receiver must constantly adjust equalizer coefficients to maintain ISI-free transmission. This is clearly not the case with the Tracker optical links, where the frequency response is fixed.

Despite the drawback of the high symbol rate, the single-carrier system avoids the need to perform a very fast FFT in the front-end. Since complexity and power consumption are the most stringent constraints in the HEP environment, a QAM system consisting of a single carrier may be more appropriate than a multi-carrier system for a future readout link.

**Table 1:** Summary of advantages and disadvantages of the single- and multi-carrier QAM systems.

|   | **Single Carrier** | **Multi-Carrier** |
|---|---|---|
| **+** | <ul><li>No need to perform an FFT in the front-end</li><li>One carrier at ~5bits/symbol – less amplitude/phase resolution needed.</li><li>PAPR is an order of magnitude less.</li></ul> | <ul><li>Simpler equalization required at the receiver.</li><li>Lower symbol rates are used (lower clock speeds required).</li><li>Data rate is more adaptive, depending on the noise characteristics of the channel.</li></ul> |
| **-** | <ul><li>Fast (possibly adaptive) equalization required at the receiver.</li><li>Huge symbol rate – more demanding on the front-end electronics.</li><li>With one carrier and a huge symbol rate, it is not easy to change data rate incrementally.</li></ul> | <ul><li>More complexity in the front-end to synthesize carriers (very fast FFT)</li><li>Some carriers will need to be loaded with as many as 8-10bit/symbol. Higher resolution needed.</li><li>Huge PAPR – transmission power backoff required</li></ul> |

## 2.4 Hardware Complexity

An upgrade will require additional components in the front-end. Digitization will have to be performed on the front-end hybrids and components for single or multi-carrier QAM modulation will also be required. These must meet the same strict requirements as the current front-end components. In this section, a single-carrier implementation is examined and the reader should refer to Figure 4 and Figure 6 for the basic mixed-signal implementation of a single-carrier QAM modulator.

Assuming the symbol rate is 640MS/s and 32-QAM is used (5bits/symbol), the total data rate would be 3.2Gbit/s. The 3.2Gbit/s bitstream from the detectors is grouped in 5-bit words. The I/Q mapper translates these into in-phase and quadrature values and must operate at the symbol rate of 640MS/s. If pulse shaping is used, a certain amount of oversampling is required. With ×4 oversampling, the sample rate at the output of the digital filters would be 2.56GS/s, which requires very high-speed DACs to be used prior to the analog stage. Even if no pulse shaping is used, the DACs must operate at 640MS/s. The analog portion of the modulator would require RF components with a modulation bandwidth of over 640MHz.

It is clear that this upgrade path represents a big challenge in terms of the effort required to overcome technological obstacles and develop a robust readout system. QAM modulators capable of Gbit/s data rates are not currently in use by any other standardized application. The implication of this is that a custom ASIC employing a proprietary modulation scheme would have to be built. This undoubtedly requires a large amount of R&D effort, possibly in collaboration with an industrial partner or experienced academic departments. It is estimated that a QAM modulator ASIC alone requires roughly 1 man-year to be produced. Including FEC and equalizer block designs, system development and testing, the total effort required to implement the QAM-based upgrade could be roughly estimated to be at least 5 man-years. Moreover, this would represent a departure from the conventional wisdom in developing optical links for HEP, which normally involves the use of commercial off-the-shelf (COTS) components with minimal customization. Hence the proposed upgrade cannot *currently* take advantage of the cost-effective approach of following industrial trends.

Bandwidth efficient modulation over fiber for high-speed links has recently attracted interest by telecoms researchers [10-13]. A future trend toward more complex modulation schemes to increase telecom optical link throughput could potentially benefit the feasibility of the proposed upgrade path. One promising approach is the one proposed by Azadet and Saibi of Agere Systems and is described in section 3.

## 2.5 Radiation Hardness

Radiation hardness of the current front-end optoelectronic components has been verified for LHC operation during the system's development and qualification [14-16]. The SLHC will present an even more harsh radiation environment, and the study of the effect this will have on laser transmitters has been initiated in [17].

A future link will also include a radiation-hard ASIC to perform the QAM modulation. The radiation effects on electronics that are of interest are total dose effects and Single Event Upsets (SEUs). Layout techniques such as enclosed layout transistors can be used to mitigate the former [18]. ASICs can be made SEU-tolerant mainly by architectural and circuit techniques such as triple modular redundancy (majority voting) logic circuits [19]. In any case, and even if a commercial QAM modulator ASIC were available, it would need customization to withstand the radiation environment of a HEP experiment.

## 2.5 Power Consumption

Power consumption is a major constraint in the CMS Tracker environment and therefore an important criterion of the upgrade's feasibility. It is difficult to make an accurate estimate of the power dissipation that can be expected in a future modulator ASIC, given that this is dependent on the technology, features and performance, as well as the particular implementation.

Agere Systems have developed a QAM transceiver which will be described in the next section. It offers performance and characteristics similar to those required for the proposed QAM upgrade and can therefore be used to obtain a 'ballpark' figure. The 1.5V, 0.14μm CMOS Agere ASIC achieves 340mW for a single-carrier 16-QAM system operating at 666MBaud (2.664Gbit/s). While FEC –which would add to the power consumption– is not included in the chip, it incorporates both a modulator and a demodulator (whereas the proposed Tracker upgrade assumes a unidirectional link). Hence, with current technology, one could expect power consumption of the order of a few hundred mW at the transmitting end of a future QAM-based link.

## 3 The Example of the Agere Systems QAM Transceiver for Optical Fiber

Currently there are no commercial applications employing QAM modulation at data rates that are sufficiently high for a future HEP readout link. Instead, high speed links use simple modulation schemes at the expense of bandwidth which is abundant in state-of-the-art optical fiber-based systems. Hence there are no Gbit/s QAM modems available commercially that could be used as a paradigm for the Tracker upgrade. However, bandwidth efficient RF modulation over fiber has recently attracted the attention of researchers as demonstrated by the example of this section. Azadet and Saibi of Agere Systems have proposed a QAM system for 40Gbit/s optical links [12, 13]. Their approach is very similar to the proposed CMS Tracker optical link upgrade and is included here to give an idea of the complexity and hardware required for implementation.

The Agere proposal involves Sub-Carrier Multiplexing (SCM) of 16 16-QAM carriers, each with a symbol rate of 666MS/s. SCM differs from Orthogonal Frequency Division Multiplexing (OFDM) in that the carrier spacing is not equal to the symbol rate. Instead, the Agere proposal employs 833MHz spacing which avoids the stringent requirements of OFDM systems for mitigating inter-carrier interference. This system is designed for an optical link with a much higher bandwidth (the last carrier is placed at ~13.3GHz). However, the idea is relevant since the symbol rate of each *individual* carrier is compatible with the frequency range available in the Tracker optical links (~0-700MHz). The motivation behind using bandwidth efficient modulation is to produce a more cost-effective system, by employing signalling rates that allow implementation in conventional IC technology such as SiGe or CMOS (transmission speeds of over 10Gbit/s normally require more expensive IC technologies).

To prove the viability of the concept, Agere have implemented a single-carrier 16-QAM transceiver operating at a symbol rate of 666MS/s [13], in a 1.5V, 0.14μm CMOS ASIC (Figure 10). The total area is 3.6mm$^2$, and the power consumption is 340mW, which includes modulation, demodulation and pseudo-random data generation on chip. This is roughly six times higher than the consumption of the Tracker AOH transmitter (which operates at 40MS/s), though one would expect that lower power consumption could be achieved in a future implementation.

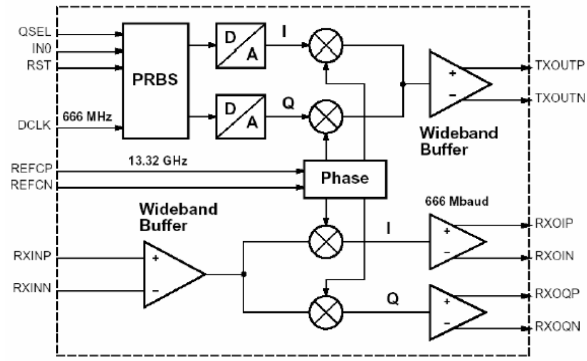

**Figure 10:** Block diagram of the Agere QAM transceiver ASIC [12].

Agere's implementation proves that the concept of QAM modulation over optical links at Gbit/s rates is technologically feasible, though not directly applicable to a HEP readout link yet. A future, lower-power ASIC which includes FEC would be necessary to demonstrate the feasibility of the Tracker upgrade, but will only materialize commercially if the telecoms industry finds sufficient motivation to adopt the concept (possibly for low-cost, multi-Gbit/s optical links).

# 4   Conclusions and Future Prospects

It has been shown that a bandwidth efficient modulation scheme could be used in a future upgrade of the CMS Tracker optical links for operation in SLHC. Laboratory tests [2, 3] have provided experimental proof of the concept, while the technological feasibility of building a QAM modulator with sufficient performance and reasonable power consumption, has been qualitatively demonstrated in this note.

The Agere ASIC proves that it is possible to produce a relatively low-power, high-speed QAM modulator. The transceiver dissipates ~0.12nW/bit/s, compared to ~0.2nW/bit/s for the current analog PAM transmission system[4]. Developing a similar chip at CERN would require approximately 1 man-year of design effort and we estimate that a full system might require in excess of 5 man-years.

The future Tracker optical links will include some form of FEC if sufficient performance is to be achieved at the Gbit/s data rates envisaged. For it to be beneficial, encoding needs to be tailored to the type of modulation chosen. Hence there is research, which was not covered in this note, to be carried out in this area.

The CMS Tracker sub-detector will have to be replaced for operation in the SLHC. Hence, it is not known how many channels will have to be read out in the future, and at what data rate. Assuming the installed optical fibers will remain in place regardless of the future implementation, it remains to be seen if 40 000 optical links transmitting data at 3-4Gbit/s will be sufficient for the readout system. The QAM-based upgrade may therefore be marginal in terms of the throughput it can sustain. Moreover, it solves only the upstream (from the detectors to the counting room) data transfer bottleneck and does not address the need for an improved downstream path for a future Timing and Trigger Control (TTC) system.

---

[4] This is based on the *equivalent* digital data rate of the Tracker analog optical links which operate at 40MS/s with an equivalent digital resolution of 8bits (=320Mbit/s). Moreover, the Agere chip's power consumption is for *both* data transmission and reception.

If all 40 000 links are used in the future system, the power dissipation of each link would –most likely– have to remain the same as for the current version of the Tracker. The new detector readout chip that will replace the APV25 would have to consume less power to compensate for the extra power required for the QAM modulator, assuming the current laser driver chip would remain as is. With digitization and FEC necessary for a future system, this would be very difficult to achieve. Nevertheless, it is not certain how many links will be required, and this problem applies to any future digital link upgrade, regardless of whether or not optoelectronic components are retained; digitization, FEC, modulation and electro-optical conversion will have to be performed in the front-end, while keeping the power consumption at current levels.

The data rate predicted in [2, 3] is modest in comparison to more traditional approaches, with 10Gbit/s being a commercial reality today. Also considering the non-negligible R&D effort required (5 man-years), as well as the challenge of developing such a high rate QAM system, this upgrade may not be the most sensible approach currently. The preservation of optoelectronic components (transmitter, fiber, receiver) and the associated effort to select, qualify and produce them, is an obvious advantage, but this must be weighted against high and risky investment in the R&D effort for a proprietary and novel technology.

High speed serial links will be developed in any case by the HEP community, regardless of the technology selected for the CMS Tracker readout system. Given the fact that a serial link solution will most likely be adopted for the upgraded TTC system, it may be easier, less risky and more reasonable to also use the same technology in the readout path of the CMS Tracker.